\documentclass[sigconf,natbib=False]{acmart}

\usepackage{graphicx}
\usepackage{comment}
\usepackage{amsmath}
\usepackage{bbm}
\usepackage{booktabs}
\usepackage[ruled]{algorithm2e}

\AtBeginDocument{%
  }

\copyrightyear{2023} 
\acmYear{2023} 
\setcopyright{acmlicensed}\acmConference[WWW '23 Companion]{Companion Proceedings of the ACM Web Conference 2023}{April 30-May 4, 2023}{Austin, TX, USA}
\acmBooktitle{Companion Proceedings of the ACM Web Conference 2023 (WWW '23 Companion), April 30-May 4, 2023, Austin, TX, USA}
\acmPrice{15.00}
\acmDOI{10.1145/3543873.3587626}
\acmISBN{978-1-4503-9419-2/23/04}

\RequirePackage[
  datamodel=acmdatamodel,
  style=acmnumeric,
  ]{biblatex}

\begin{document}

\title{On Modeling Long-Term User Engagement from Stochastic Feedback}


\author{Guoxi Zhang}
\affiliation{%
  \institution{Graduate School of Informatics, \\ Kyoto University}
  \city{Kyoto}
  \country{Japan}
  \postcode{606-8501}
}
\email{guoxi@ml.ist.i.kyoto-u.ac.jp}

\author{Xing Yao}
\affiliation{%
  \institution{China Central Depository \& Clearing Co., Ltd.}
  \city{Beijing}
  \country{China}}
\email{854352859@qq.com}

\author{Xuanji Xiao}
\affiliation{%
  \institution{Shopee Inc.}
  \city{Shenzhen}
  \country{China}
}
\email{growj@126.com}
\authornote{Corresponding author}

\renewcommand{\shortauthors}{Zhang et al.}

\begin{abstract}
  An ultimate goal of recommender systems (RS) is to improve user engagement. Reinforcement learning (RL) is a promising paradigm for this goal, as it directly optimizes overall performance of sequential recommendation. However, many existing RL-based approaches induce huge computational overhead, because they require not only the recommended items but also all other candidate items to be stored. This paper proposes an efficient alternative that does not require the candidate items. The idea is to model the correlation between user engagement and items directly from data. Moreover, the proposed approach consider randomness in user feedback and termination behavior, which are ubiquitous for RS but rarely discussed in RL-based prior work.  With online A/B experiments on real-world RS, we confirm the efficacy of the proposed approach and the importance of modeling the two types of randomness.
\end{abstract}


\maketitle

\section{Introduction}
Recommender Systems (RS) are software systems that help users discover engaging items. In particular, a sequential RS allows users to consume new items without ceasing, usually by scrolling down the interface to the RS. The main goal of sequential RS is to increase long-term user engagement. In this regard, myopic decision-making, such as ranking items solely by predicted click-through rate (pCTR), can be suboptimal. For instance, news can have higher pCTR than dramma bloopers, yet dramma bloopers prompt more consumption for videos about the dramma and staffs in the dramma. 

Reinforcement learning (RL) has emerged as a promising research direction for optimizing long-term user engagement~\cite{10.1145/3289600.3290999,ijcai2019-360,zhao2019deep,10.1145/3178876.3185994}. Using RL, sequential recommendation is modeled as interaction between an agent and an environment. At each step, the agent receives a \emph{state} and selects an \emph{action}. Usually, states are user profiles and their historical activities, and actions are items available for recommendation. After selecting each action the agent receives a scalar \emph{reward}. The goal of RL is to learn a \emph{policy} for action selection, with which the agent can maximize the sum of rewards received during interacting with the environment. Thus, RL is an anticing paradigm if one uses utility for items (e.g. clicks) as rewards, as it naturally optimizes long-term utility of sequential RS. 

However, training RL models for industrial RS can be challenging. To understand the issue, consider a minimalist example in which an agent selects one item from a set of candidate items for a user. To train a model for pCTR, one needs to record information about the user, the selected item, and the feedbacks for items. However, to train a RL model such as the well-known deep Q-network (DQN)~\cite{mnih-dqn-2015}, one needs to additionally record the candidate item set. This induces huge overhead for industrial applications as the candidate set usually contains several hundreds of items. While the overhead can be reduced by learning RL models using item embedding vectors, it is reported that the performance is much worse than learning RL models using items directly~\cite{arXiv.2110.11073}.

This paper proposes an efficient alternative for optimizing long-term user engagement. The key observation behind the proposed approach is that in industrial setting RS are periodically trained on logs generated by some other RS, which is referred to as the behavior policy afterwards. For applications serving millions of users the behavior policy is usually highly optimized, so by mining the relation between user engagement and items selected by this policy we can obtain a model for user engagement. Notably, the proposed approach does not require the candidate item set in learning. So it can be integrated into any existing myopic RS effortlessly without introducing huge overhead.

In addition to the efficiency issue, the proposed approach also considers stochasticity in rewards and termination of interactions. RL algorithms such as DQN usually assume that (a) rewards are deterministic regarding to states and actions and (b) interactions last for infinite steps. Both of them are rarely satisfied in RS. Real-world RS need to serve users with diverse interests, but they only have limited information about users. Consider two different users that have the same feature values. As they have different interests, the feedbacks they provide for the same item will be different. From the perspective of RS, this means that the rewards for the same state and actions tend to be random. For the same reason, it is unlikely that they will interact with RS for the same number of steps. Unfortunately, both types of stochasticity have not been addressed so far. To model the former, this paper employs the recent distributional RL framework~\cite{Bellemare2017}. For the latter, this paper extends the distributional RL framework to random termination setting and learns a model to predict whether users terminate interaction.

We evaluated the proposed approach on an real-world industrial RS for short videos that serves millions of users. During a week of online A/B test, after deploying the proposed approach we were able to improves the average number of videos viewed by a user by 2.72\% and the average duration of video views by 1.59\%. Moreover, the experiment also confirmed that modeling stochasticity is essential for effectively modeling long-term user engagement. 

The contribution of this paper can be summarized as follows.
\begin{itemize}
\item This paper proposes a simple and efficient approach for modeling long-term user engagement in RS.
\item This paper proposes to consider stochasticity in user feedback and termination when modeling user engagement. 
\item This paper evaluated the proposed method on an industrial RS and confirmed its efficacy.
\end{itemize}

The rest of this paper organizes as follows. Section 2 briefly reviews  relevant literature, and section 3 provide technical background. Section 4 describes the proposed approach. Section 5 presents experiments on real-world RS, and section 6 concludes this paper.

\section{Related Work}
Existing attemts for RL-based recommendation can be classified into policy-gradient based~\cite{10.1145/3289600.3290999} and variants of the DQN algorithm~\cite{10.1145/3240323.3240374,10.1145/3178876.3185994}. This paper address the efficiency issue, which can be consider as a complementary approach for resource-constrained scenarios. 

In the literature of RS, significant effort has been devoted to neural architectures. The proposed approach leverages the DLRM~\cite{DLRM19} architecture, though it can be combined with other architectures as well. Meanwhile, to model the randomness in rewards, the proposed approach leverages the recent distributional RL framework~\cite{Bellemare2017,Dabney2018}. As for random termination, in literature it has been considered in a model-based RL~\cite{pmlr-v70-white17a} and modeled with evolutionary algorithm~\cite{yoshida2013reinforcement} or meta gradient descend~\cite{NEURIPS2018_2715518c}. Our proposed method differs from these approaches as it learns the discount factor from leaving behaviors of users.

\section{Preliminaries}
\paragraph{Modeling Sequential Recommendation with RL}
Sequential interaction between a user and a recommender agent is modeled as a discounted infinite-horizon Markov Decision Process (MDP): $<\mathcal{S}, \mathcal{A}, R, P, \gamma>$. States $\mathcal{S}$ are information about users, which include features such as interest tags and lists of items consumed in past few minutes, hours or days. Actions $\mathcal{A}$ are candidate items. In distributional RL, rewards $R$ are considered as random variables. This paper utilizes clicking signal as rewards, so a realization $r=1$ means the corresponding items is clicked, and $r=0$ otherwise. The transition dynamics $P$ governs state transitions, and the discount factor $\gamma$ is used for defining value distribution. 

\paragraph{Action Value Distribution} We are interested in modeling the the sum of rewards obtained during agent-environment interaction, which characterizes long-term user engagement. As we assume rewards to be random variables, state-action values $Z^\pi(s,a)$, the discounted sum of rewards obtained after selecting an action $a$ at some state $s$ and following $\pi$ afterwards, is also a random variable. Bellemare et al. showed that $Z^\pi(s,a)$ is the fixed point of the distributional Bellman operator $\mathcal{T}^\pi$ that is defined as~\cite{Bellemare2017} : 
\begin{equation}
\forall (s,a)\in \mathcal{S}\times\mathcal{A}, \mathcal{T}^\pi Z^\pi(s,a) \overset{D}{=} R(s,a) + \gamma Z^\pi(s', \pi(s')),
\label{dbe}
\end{equation}
where $s'\sim P(s|s,a)$. $X \overset{D}{=} Y$ means that the random variable $X$ has the same distribution as random variable $Y$. 

Dabney et al. proposed to parameterize $Z^\pi$ with the so-called quantile distributions~\cite{Dabney2018}. The idea is to learn a parametric function $\beta: \mathcal{S}\times\mathcal{A}\rightarrow \mathbb{R}^{M}$ to estimate $M$ evenly spaced quantiles of $Z^\pi$. Given data from $\pi$, they proposed to learn $Z^\pi$ by minimizing the quantile huber loss:
\begin{equation}
L(\beta)= \mathbb{E}_{(s,a,r,s')\sim\pi} \frac{1}{M}\sum_{i=1}^M \sum_{j=1}^M[\rho_{\hat \tau_i}^\kappa(r+\gamma\beta'(s',\pi(s'))_j-\beta(s,a)_i)],
\label{dpe}
\end{equation}
\noindent where $\tau_i$ is the i\textsuperscript{th} quantile of value distribution, and $\hat{\tau}_m = \frac{1}{2}(\tau_{m-1} + \tau_m)$ for $m=1,2,\dots, M$ are quantile mid-points. For simplicity, $\tau_0=0$. $\rho_\tau^\kappa(u) = \frac{1}{\kappa}|\tau-\mathbb{I}(u<0)|\rho_\kappa(u)$ and $\mathbb{I}(\cdot)$ is the indicator function. $\kappa>0$ is a hyper-parameter, and in practice we use $\kappa=1$. $\beta'$ has the same neural structures as $\beta$, and its parameters are periodically copied from $\beta$.

\section{Proposed Method}
\subsection{The Learning Problem}
We assume an agent learns from some offline data $\mathcal{D}$ that are generated by behavior policy $b$. $\mathcal{D}$ are logs of sequential interactions between users and the agent. We represent the information about the user and as states and recommended items as actions. The reward for a decision equals to one if the corresponding item is clicked, and it equals to zero otherwise. Finally, we organize $\mathcal{D}$ as transitions, which can be written as $(s,a,r,s')$, where $s$ and $s'$ are two consecutive states, $a$ is the action chosen at $s$ and $r$ is the reward for this decision.

Our goal is to learn a function $g(s,a):\mathcal{S}\times\mathcal{A}\to\mathbb{R}$ that predicts the the sum of rewards obtained after selecting $a$ at step $s$ and following $b$ afterwards, which characterizes the long-term user engagement of selecting $a$. With $g(s,a)$, we can rank items according to their long-term user engagement.

\subsection{The Proposed Approach}
The proposed method is based on the following observation. Real-world RS are often highly optimized using both learning-based method and rule-based methods, and they are regularly updated to adapt to new users and new items. As a result, users tend to interact with them sequentially, so the data generated by them contain information about items' effect on user engagement. Thus, we propose to mine such information from data.

Meanwhile, as discussed in Section 1, user feedbacks are stochastic by nature. In consequence, when modeling user engagement as cumulative user feedbacks, we have to appropriately address such randomness. The idea of the proposed approach is to estimate $Z^b(s,a)$, the value distribution of behavior policy $b$ from data $\mathcal{D}$. This can be achieved with Equation~\ref{dpe}. With $Z^b(s,a)$ we can use its expectation as a scoring function for long-term user engagement, i.e. $g(s,a)=\mathbb{E}[Z^b(s,a)]= \frac{1}{M}\sum_j^M\beta(s,a)_j$.

We now discuss how to handle random termination. While the discount factor $\gamma$ in an MDP is used to defined value functions and value distributions, it can also be interpreted as the probability that interaction continues after a state-action pair~\cite{Littman1994}. Thus, $1-\gamma$ is precisely the probability that interaction terminates. Our idea is to estimate such probability using data and learn $Z^b(s,a)$ based on such estimate. In consequence, $Z^b(s,a)$ will be aware of the correlation between decisions and termination of interactions. This leads to a new distributional Bellman operator $\mathcal{T}^{\pi}_{\ell}$:
\begin{equation}
\forall (s,a)\in \mathcal{S}\times\mathcal{A}, \mathcal{T}^{\pi}_{\ell} Z^\pi(s,a) \overset{D}{=}  R(s,a) + \gamma'(s',\pi(s'))Z^\pi(s', \pi(s')),
\label{leaving}
\end{equation}
\noindent where $\gamma'(s,a) := \min(1-\ell(s,a), \eta)$. $\ell(s,a)$ is the probability that interaction terminates after $(s,a)$, and $\eta\in(0,1)$ is a hyper-parameter. In this operator, the value distribution of next state $Z^\pi(s',\pi(s'))$ is weighted by $\gamma'(s,a)$, which characterizes the probability that interaction continues after $(s', \pi(s'))$. Thus, with this operator the learned $Z^b(s,a)$ will be able to model random termination. 

One potential issue about $\mathcal{T}^{\pi}_{\ell}$ is its convergence. In what follows we show that $\mathcal{T}^{\pi}_{\ell}$ is a contraction in tabular case, based on the analysis for $\mathcal{T}^{\pi}$ provided by~\cite{Bellemare2017}. To begin with, let $F_X$ and $F_Y$ be cumulative distribution functions (CDF) of random variable $X$ and $Y$. The Wasserstein metric $d_p$ between $F_X$ and $F_Y$ can be written as $d_p(F_X, F_Y)=\|F_X(u) - F_Y(u) \|_p$\footnote{To simplify notation, random variables and their CDF are conflated whenever possible.}. For two action value distributions $Z_1$ and $Z_2$, defined $\bar{d}_p(Z_1, Z_2) := \sup_{s,a} d_p(Z_1(s,a), Z_2(s,a))$.
Our analysis is based on following two properties of $d_p$~\cite{Bellemare2017}. Suppose $A$ is a random variable that is independent with $X$ and $Y$ and $a$ is a constant, then:
\begin{equation}
\begin{split}
d_p(A+X, A+Y)&\leq d_p(X,Y), \\
d_p(aX, aY) &\leq |a| d_p(X, Y). \\
\end{split}
\end{equation}

Suppose we have samples from policy $\pi$. Then for two distributions over $\mathcal{S}\times\mathcal{A}$, $Z_1$ and $Z_2$, and for any state $s\in\mathcal{S}$ and $a\in\mathcal{A}$, we can characterize the Wasserstein metric between after applying $\mathcal{T}_\ell^\pi$ as:
\begin{equation}
\begin{split}
d_p(\mathcal{T}_\ell^\pi Z_1,& \mathcal{T}_\ell^\pi Z_2)\\
& = d_p(R(s,a) + \gamma' Z_1(s', \pi(s'), R(s,a) + \gamma' Z_2(s', \pi(s')) \\ 
&\leq d_p(\gamma'(s',  \pi(s'))Z_1(s',  \pi(s')), \gamma'(s',  \pi(s'))Z_2(s',  \pi(s'))) \\
& \leq \eta d_p(Z_1(s',  \pi(s')),Z_2(s',  \pi(s'))) \\
& \leq \eta \mathrm{sup}_{s', a'}  d_p(Z_1(s',  a')),Z_2(s',  a')) \\
& = \eta \bar{d}_p(Z_1, Z_2) \\
\end{split}
\end{equation}
Hence $\bar{d}_p(\mathcal{T}_\ell^\pi Z_1, \mathcal{T}_\ell^\pi Z_2)=\mathrm{sup}_{s,a}d_p(\mathcal{T}_\ell^\pi Z_1, \mathcal{T}_\ell^\pi Z_2)\leq \eta \bar{d}_p(Z_1, Z_2)$, which means $\mathcal{T}^\pi_{d,\ell}$ is a $\eta$-contraction in $\bar{d}_p$.

\subsection{Practical Algorithm}
As for a practical algorithm, we estimate $Z^b(s,a)$ and $\ell(s,a)$ simultaneously using data $\mathcal{D}$. The objective function to be minimized is as follows. 
\begin{equation}
  \begin{split}
  &L_{\ell}(\beta,\ell)=\\
  &\frac{1}{|\mathcal{D}|}\sum_{(s,a,r,s')\in\mathcal{D}}\{ \frac{1}{M}\sum_{i=1,j=1}^M[\rho_{\hat \tau_i}^\kappa(r+\gamma'\beta'(s',\pi(s'))_j-\beta(s,a)_i)] \\
  &- e\log \ell(s,a) - (1-e)\log (1-\ell(s,a)) \}.
  \end{split}
  \label{eq:obj}
\end{equation}
\noindent $e=1$ if interaction terminates at $(s,a)$ and $e=0$ otherwise. $\beta$ and $\ell$ can be parameterized as a multi-objective network, in which they share feature embeddings but have different feedforward networks. Algorithm~\ref{iltum} provides pseudo code for this algorithm.

\begin{algorithm}[t]
\caption{A practical algorithm for the proposed framework)}
\KwIn{$T_c$, frequency for copying $\beta$ to $\beta'$\newline
$B$, the size of mini-batches}
Initialize $\beta$ randomly and copy its values to $\beta'$ \\
\For{mini-batch $i=1, 2,\dots, \frac{n}{B}$}{
Update $\beta$ and $\ell$ using Equation~\ref{eq:obj}.\\
\If{$i \mod T_c = 0$}{ $\beta'=\beta$}
}
\label{iltum}
\end{algorithm}

\section{Experiments}
This section evaluates the proposed method on an industrial RS that generates a feed for ''videos you might also like''. In particular, we investigates the following two questions.
\begin{enumerate}
    \item Can we improve user engagement of real-world RS via modeling $Z^b(s,a)$?
    \item Does modeling randomness in rewards and termination beneficial in practice?
\end{enumerate}

\subsection{Methodology}
We followed standard protocol for A/B testing. Users were randomly hashed into the control group and test groups. Before the evaluation period, we confirmed that there was no significant difference between the control group and test groups using statistical testing. Every group has at least 630,000 users. 

As the our goal is to improve RS via modeling long-term user engagement, we evaluated the performance of the RS after deploying algorithms to the system. Let $g^b(s,a)$ be the current ranking function of the RS. Then after deploying estimator for user engagement $g(s,a)$, the new ranking function becomes $g^b_\text{new}(s,a) = g^b(s,a) + w * g(s,a)$, where $w$ is a parameter tuned in preliminaries experiments.

We used the metrics for video consumption as evaluation metric for user engagement, which are listed in Table~\ref{metrics}. We report the change in performance after deploying algorithms to the RS. A positive change indicates that an algorithm can improve the performance of the current online policy.

\begin{table}[t!]
\caption{Evaluation metrics used in online experiments. VV is the number of videos clicked by a user, and IMP is the number of videos shown to a user. DUR is the sum of time a user spent on watching videos. These three metrics measure the overall performance during the A/B test, so we consider them as metrics for user engagement. CTR and CVR are metrics for individual videos. They reflect the extent to which a user likes individual items, so they are metrics for short-term utility.}
\label{metrics}
\centering
\begin{tabular}{@{}ll@{}}
\toprule
Metric              & Meaning                                                 \\ \midrule
video views (VV)   & \#clicks / \#user                                  \\
\#impression (IMP) & \#impression / \#user                              \\
duration (DUR)     & $\sum$(play duration) / \#user                        \\
CTR                & \#clicks / \#impression                            \\
CVR                & $\sum$(play duration) / $\sum$(clicked videos length) \\ \bottomrule
\end{tabular}
\end{table}

\subsection{Algorithms}
\paragraph{control group} A highly-optimized RS for short-videos used in some industrial application.
\paragraph{Proposed} The proposed approach, which considers randomness in rewards and termination for modeling user engagement.
\paragraph{Random-Reward} A variant of the proposed approach considers randomness in rewards but not randomness in termination. Refer to this method as RR.
\paragraph{Full-Deterministic} A variant of the proposed approach that directly estimates $g(s,a)$ without considering randomness in rewards and termination. Refer to this method as FD.

\subsection{Implementation Details}
We used the DLRM structure in experiments. There were 40 features in total, including user ID, item ID, and item keywords. These discrete features were mapped to embedding vectors in $\mathbb{R}^{32}$. Parameters of the target module are copied from the policy module every 100 gradient steps. Algorithms were trained daily using a single P40 GPU. We used Adam as the optimizer for all three algorithms with a learning rate set to $0.00015$. Whem modeling randomness in rewards, the number of quantiles was set to 200. 

\begin{table}[t]
  \caption{Results of our online A/B test. The numbers shown in parentheses are p values. Without considering any randomness, FD failed to improve any metric significantly. After considering randomness in rewards, RR was able to improve VV, CTR and CVR significantly. The proposed apporach improved VV, DUR, CTR and CVR significantly. The results confirm that modeling randomness in rewards and termination is crucial for modeling user engagement.}
  \label{online_eval}
  \centering
  \begin{tabular}{@{}llll@{}}
  \toprule
  Metrics  & FD                & RR                & Proposed                      \\ \midrule
  VV            & 1.19\% (0.11)  & \textbf{1.58\% (0.031)}  & \textbf{2.72\% ($2.1\times 10^{-4}$)} \\
  IMP           & 0.140\% (0.86) & -0.265\% (0.74)          & 0.502\% (0.53)           \\
  DUR           & 0.263\% (0.73) & 1.24\% (0.10)            & \textbf{1.59\% ($3.5\times 10^{-2}$)}  \\
  CTR           & 1.05\% (0.93)  & \textbf{1.85\% ($3.6\times 10^{-3}$)} & \textbf{2.21\% ($4.1\times 10^{-4}$)} \\
  CVR           & 0.410\% (0.14) & \textbf{1.22\% ($3.0\times 10^{-6})$}     & \textbf{1.07\% ($4.8\times 10^{-5}$)} \\ \bottomrule
  \end{tabular}
  \end{table}
\subsection{Results}

Table~\ref{online_eval} shows results for our online evaluations. FD failed to provide any significant improvement over the control group. RR was able to improve VV, CTR and CVR significantly, though its improvement for DUR was not significant. These results imply that users in this group were likely to watch slightly more videos than the control group. Moreover, they show that modeling randomness in rewards is indeed helpful for characterizing items' effect on user engagement. Meanwhile, the proposed method was able to significantly improve VV, DUR, CTR and CVR, showing that randomness in termination is yet another key factor for modeling user engagement. Together, Table~\ref{online_eval} confirms the efficacy of the proposed method as an efficient alternative for modeling long-term user engagement for RS.

\section{Conclusion}

This paper investigates how to model long-term user engagement for RS efficiently. The proposed approach relies on an observation that the behavior policy (i.e. the RS in production environment) in industrial applications are often highly optimized. So instead of learning the optimal policy from scratch, the proposed model tries to capture correlation between user engagement and recommended item in offline data. Although not guaranteed to be optimal, it enjoys computational efficiency and is confirmed to be effective for real-world systems. Moreover, this paper proposes to model randomness in rewards and termination and confirms the importance of the two factors for modeling user engagement.

\printbibliography

\end{document}